\documentstyle[prb,aps,epsf,twocolumn]{revtex}
\draft

\begin{document}
\title{Does the melting behavior in ultrathin metallic nanowires begin from the surface ?}

\author{Jinlan Wang$^{1,2}$, Xiaoshuang Chen$^2$, Guanghou Wang$^1$\thanks{To whom the correspondence should be addressed. E-mail: wangqun@nju.edu.cn}} \author{Baolin Wang$^1$, Wei Lu$^2$ and Jijun Zhao$^3$} 

\address{$^1${\it National Laboratory of Solid State Microstructures and Department of Physics Nanjin University Nanjing 210093, P.R. China}\\ 
$^2${\it National Laboratory for Infrared Physics, Shanghai Institute of Technical Physics, Chinese Academy of Sciences, Shanghai 200083, P.R. China}\\
$^3${\it Department of Physics and Astronomy, University of North Carolina at Chapel Hill, North Carolina 27599-3255}}
\maketitle

\begin{abstract}

Surface melting temperature is well-known significantly lower than the bulk melting point. But we find that the interior melting temperature in ultrathin nanowires is lower than that of the surface melting . The thermal stability of helical multi-walled cylindrical gold nanowires is studied using molecular dynamics simulations. The melting temperature of gold nanowires is lower than the bulk value, but higher than that of gold nanoclusters. An interesting interior melting is revealed in the gold nanowires and the thermodynamics is closely related to interior structures. The melting starts from the interior atoms, while the surface melting occurs at relatively higher temperature. We propose that the surface melting represents the overall melting in ultrathin metallic nanowires. 
\end{abstract}

The melting behavior of nanoparticles and nanorods have been demonstrated dramatically different from the bulk both experimentally and theoretically \cite{labasite,ercolessi,berry,gulseren,goldstein,wangzl,link,schmidt,nayak,hu,liu,lee,wu}. The melting process of a crystalline starts from the surface layer and propagates into the interior. Thus, the surface melting temperature is significantly lower than the bulk melting point. Similarly,  people may ask whether the surface melting temperature is lower than the overall melting temperature in clusters and nanowires. Berry considers that ``dynamic coexistence'' or surface melting happens in the melting process of small clusters before the total melting\cite{berry}. For crystalline nanowires, Tosatti found that surface melting temperature for Pb wires was also lower than the total melting temperature\cite{gulseren}. Experimentally, surface melting is involved in the melting process of nanoparticles and nanorods \cite{goldstein,wangzl,link}. Especially, Schmidt found a broad peak for heat capacity of Na$_{139}^{+}$ clusters, maybe implying the coexistence of solid-like and liquid-like phases before the total melting \cite{schmidt}. Two major effects are responsible for these different melting behavior in nanoparticles and nanorods. One is the large surface-to-volume ratio in these nanostructures. The other is quantum confinement effect in finite size systems. Surface atoms have fewer nearest neighbors and weaker binding, which may induce an earlier surface melting behavior. On the other hand, a close relationship between the melting and the structure features is found in the cluster\cite{kusche,litx}. 

Recent studies demonstrated that ultrathin metallic nanowires have quite different structural and properties from those of bulk, clusters and crystalline nanowires \cite{kondo,kondo1,tosatti,wang,bilalb1,bilalb2,bao,bao1}. The helical multi-walled cylindrical structures have been widely found in 1-3nm size range of metallic nanowires both experimentally and theoretically \cite{kondo,kondo1,tosatti,wang,bao,bao1}. This kind of novel structure will bring some bizarre melting features different from the above-mentioned systems. To our knowledge, less efforts are focused on their thermodynamics so far, although this kind of ultrathin metallic nanowires have attracted great interests \cite{bilalb1,bilalb2}.

Furthermore, the ultrathin nanowire has some characteristics similar to the cluster, crystalline wire and bulk. It may provide an opportunity for comprehensively understanding these types of matter states and their relations. For example, the surface-to-volume ratio is a non-zero value in these nanostructures, while it approaches toward zero in the bulk. Surface and core atoms are expected to play different role during the melting process. However, it is difficult to definitely distinguish the overall melting and surface melting of clusters experimentally because the cluster's signal spectrum will not exist when the surface melting occurs. Therefore, we can employ the ultrathin gold nanowire as a representation to explore these problems.

In this letter, the thermal stabilities of gold nanowires with helical multi-walled cylindrical structures are studied by using molecular dynamical (MD) simulations. We start from the optimized structures from previous works \cite{wang}, which were imaged by electron microscopy\cite{kondo1}. The interaction between gold atoms is described by a glue potential \cite{glue} and the periodic boundary condition is applied along the wire axis direction to model the wire with sufficient length wire. The length of supercell is chosen with the same as Ref. \cite{wang}, which is a reasonable scheme to attain the helical structures in the nanowire. To characterize the thermal behavior of nanowires, we monitor the root-mean-square (rms) fluctuation of the interatomic bond distances $\delta $ defined by 

\begin{equation}
\delta =\frac 2{n(n-1)}\sum\limits_{i<j}^n\frac{\left( \left\langle
r_{ij}^2\right\rangle _t-\left\langle r_{ij}\right\rangle _t^2\right) ^{1/2} 
}{\left\langle r_{ij}\right\rangle _t}
\end{equation}

and the heat capacity $C$ per atom, which is related to the energy $E$ fluctuation, by the relation

\begin{equation}
C=\frac{(\langle E^2\rangle -\langle E\rangle ^2)}{nk_BT^2}
\end{equation}

where $r(i,j)$ denotes the distance between the nuclei $i$ and $j$, $n$ is the total number of the atoms in the nanowire, $k_B$ is the Boltzman constant, $\langle ...\rangle $ indicates the thermal statistical averages in the canonical ensemble after equilibration. 

The constant temperature molecular dynamics (MD) method of Nose \cite{nose} is employed to exploit the thermal properties of gold nanowires. The MD time step is chosen as 2.15$\times 10^{-15}$s. The initial 10$^5$
steps are used to bring the system into equilibration and further 10$^6$ steps are used to record the thermal average of the physical quantities. We study five representative helical structures of gold nanowires with up to
triple shells: single atom centered, double-chain, trigonal, tetragonal and six atoms parallelogram centered noted as S1, S2, S3, S4, S6 in Fig.1. Table I presents the starting melting temperature $T_{ini}$ and the overall
melting temperature $T_m$ of these structures. The overall melting temperature $T_m$ reflects the stability of the whole system, while the starting melting temperature $T_{ini}$ describes the early stage of melting
and sensitively depends on wire structures. The overall melting temperature $T_m$ is estimated from the curves of rms bond length fluctuation $\delta $, heat capacity $C$ and binding energy $E$ as functions of temperature.

\begin{figure}
\centerline{
\epsfxsize=3.5in \epsfbox{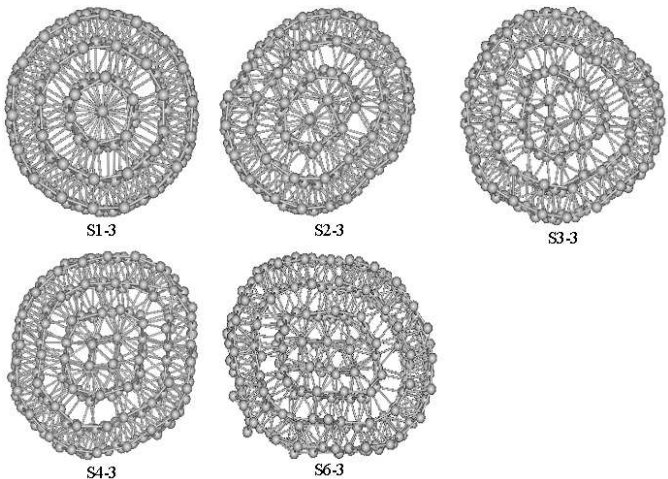}
}
\caption{Morphology of some representative gold nanowires with triple shells.}
\end{figure}

As shown in Table I, the overall melting temperatures $T_m$ for nanowires are almost the same for all the wires and are slightly lower than the bulk melting point. Experimentally, Liu {\em et al. } have reported the melting point of Pt nanowire is about 400$^o$C\cite{liu}. Lee {\em et al.} have also found the melting temperature of 4.6nm Pd nanowire is just 300$^o$C, which is much lower than the bulk value (1445$^o$C)\cite{lee}. Similar phenomenon is also found in the case of metal clusters. The depression of melting point can be attributed to the low dimension and large surface-to-volume ratio in these nanostructures. However, the melting temperatures obtained for gold nanowires are much higher than those of gold nanoclusters \cite{Garz,Buffat,Castro,Jellinek}. This may be understood by the tightened helical structures in nanowires. On the other hand, the starting melting temperatures $T_{ini}$ are different for wires with different interior structures (see Table I). The wire S1 (as described by multi-walled structural index:18-12-6-1) starts to melt at a rather lower temperature than the other ones, while the wire S4 (21-15-9-4) has relatively higher starting melting temperature. These indicate that the different starting temperature $T_{ini}$ for different nanowires are related to their different interior structures. Similar overall stability for all the wires may come from their common multi-walled helical packing, despite different interior structure. Moreover, since the $T_{ini}$ is related to the stability of interior atomss, the rather lower starting melting temperature $T_{ini}$ implies that the interior melting behavior happens during melting process of the helical gold nanowires.

We discuss the structural evolution of gold nanowires during the melting process. Fig.2 gives several snap shots taken from the structural trajectories of the wire S1 at different temperatures. It is interesting to note that the interior atoms diffuse along the wire axis direction at a rather low temperature. As shown in Fig.2, the center atoms firstly move along the wire at 300K (Fig.2a). The helical structure of the outmost shell is almost invariant. With the raise of temperature, the center atoms continue moving away from the wire. The atoms in the first shell (from interior to outer) thus have fewer nearest neighbors and begin to diffuse along axis direction (Fig.2b). Similarly, the atoms in the second shell are also involved into the migration at higher temperature $T=900$K (Fig.2c). However, the helical structure of surface shell still exists. When the temperature is high enough, the surface atoms in the outmost shell also start to migrate and the helical structure of surface is broken at 1100K, leading to the overall melting (Fig.2d). Therefore, we conclude that interior atoms diffuse prior to surface atoms and no surface melting takes place before the overall melting in gold nanowires.

\begin{table}[tbp]
Table I. Melting temperature for different interior structure of Au nanowire.
\par
\begin{center}
\begin{tabular}{ccccccc}
melting temperature(K) & S1 & S2 & S3 & S4 & S6 &  \\ \hline
T$_{ini}$ & 300 & 550 & 650 & 700 & 650 &  \\ 
T$_{m}$ & 1100 & 1100 & 1050 & 1100 & 1100 & 
\end{tabular}
\end{center}
\end{table}

\begin{figure}
\centerline{
\epsfxsize=3.5in \epsfbox{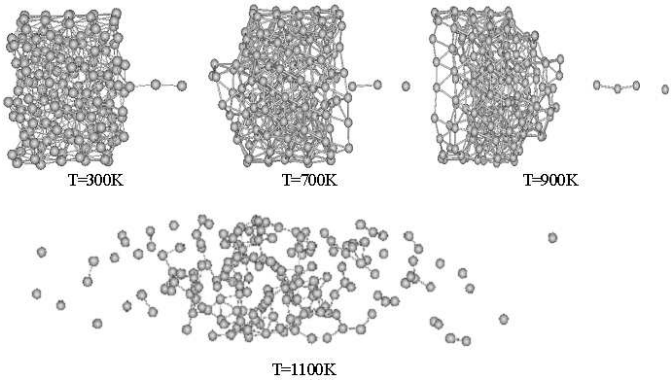}
}
\caption{Structure evolution of gold nanowire S1(18-12-6-1) with elevated temperature.}
\end{figure}

To further illustrate the argument of the interior melting and distinguish the role of surface and core atoms on the melting behavior, Fig.3 plots the rms bond length fluctuation of the surface atoms ($\delta _s$), core atoms ($\delta _c$) and the total atoms ($\delta $) of S1 wire as functions of temperature. Obviously, the rms bond length fluctuation of the core atoms have similar trends to that of the whole wire, but dramatic different from
the surface. In the temperature range of $350-1000$K, the $\delta _s$ of the surface atoms is very small and almost invariable, while $\delta _c$ and $\delta $ have substantial fluctuations. In other words, the core atoms begin to diffuse along axis and become `wet', while the surface atoms remain `solid-like' at this melting region. These results indicate that the melting mainly comes from the diffusion of core atoms and no surface melting happens at the beginning of the melting. Moreover, the $\delta _c$ fluctuates around 0.12 at low temperature, consistent with the Lindemann criterion for equilibrium melting of simple crystals\cite{11}, which shows that the interior melting takes place. For all the three cases, there is a rapid raise of rms bond length fluctuation in a narrow temperature region (1000-1150K), indicating that the surface atoms are involved into the melting process. Afterwards, the surface atoms play an important role on the melting of the wire at the high temperature region. Above 1150K, all of three quantities have a large, constant and smooth variation, corresponding to the completely melting status. Together with Fig.2d, we hold the surface melting means the overall melting in the multi-walled ultrathin helical nanowires. The interior melting is in preference to the surface melting and the diffusion of core atoms has a dominant effect on the melting at low temperature. The surface atoms affect the melting at high temperature and surface melting represents the overall melting.

The above interior melting behavior is obviously different from that of the bulk, clusters and crystalline nanowires, where the surface melting usually occurs before the total melting. In the bulk, after the melting of surface layers, the rest core atoms can still be seen as an analogous bulk. Much more energy is needed to make the `rest bulk' molten. Therefore, the surface melting temperature can be lower than the overall melting point. In the cases of clusters and crystalline nanowires, surface atoms have fewer nearest neighbors and are thus weaker bound and less constrained compared to the core atoms. Thus, surface atoms are easy to diffuse and become liquid-like. However, the present structures of gold nanowires are helical multi-walled cylindrical. The interaction among the same shell atoms is stronger than that of neighboring shell atoms. Furthermore, to reach a well-tightened helical multi-walled structure, the helical match may cause the fewer number of atoms in the interior shells, especially for the center chain. Thus, interior atoms may have less coordinate numbers and larger interatomic distances than those of surface atoms. Therefore, core atoms in nanowires can break away from the binding sites in the wire prior to surface atoms and diffuse at a rather low temperature. In addition, in th  optimization process of ground state structure, we have found that the formation of helical structure in the outer shell is earlier than the interior shell. This implies that the surface is dynamically more stable than the interior part. Experimentally, Wu {\em et al.} have found that the melting of Ge nanowire starts from the two ends of the wire and move towards the middle\cite{wu}. These imply that the melting behavior for nanowires dramatically differs from those for nanorods and bulk\cite{link}.

\begin{figure}
\vspace{0.5in}
\centerline{
\epsfxsize=3.3in \epsfbox{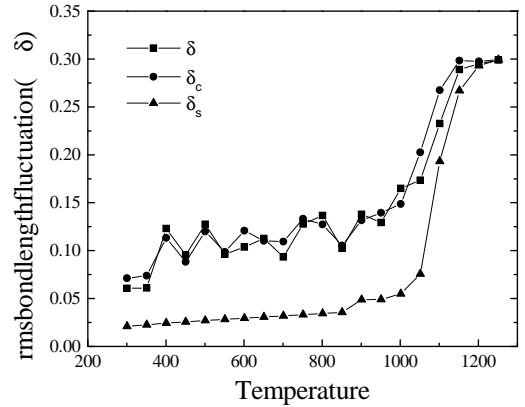}
}
\vspace{-0.7in}
\caption{The rms bond length fluctuation in the S1(18-12-6-1) structure is ploted as functions of temperature for the entire system($\delta $), the core atoms($\delta _c$) and the surface atoms($\delta _s$).}
\end{figure}

To further clarify our ideas, we separate the function of surface atoms and core atoms by fixing the surface atoms artificially and allowing the interior atoms to move or vice verse. All the rms bond length fluctuation $\delta $, $\delta _{fc}$, $\delta _{fs}$ are calculated for the entire system, representing no constrained, fixing core atoms and fixing surface, respectively. We still take the wire S1 as an example. As shown in Fig.4, regardless the large difference among $\delta $, $\delta _{fc}$, $\delta_{fs}$ curves, the full melting temperatures (which corresponds to a large, smooth and constant rms bond length fluctuation) in the three cases are about 1150K. The $\delta _{fs}$ has considerable jump at low temperature (400K), while the $\delta _{fc}$ is almost invariable up to 800K. This indicates that surface atoms with the helical structure are thermodynamically more stable. For the case of fixing the surface and no constrained, although the $\delta_{fs}$ is just nearly half of the $\delta $, their general trends are similar. The small absolute value of $\delta_{fs}$ comes from the fixing surface atoms, which contributes nearly zero to the rms bond length fluctuation. Therefore, we propose there should be no essential difference between the first two cases. These results also support the melting comes from the interior atoms at low temperature and the surface melting represents the overall melting.

\begin{figure}
\vspace{0.5in}
\centerline{
\epsfxsize=3.3in \epsfbox{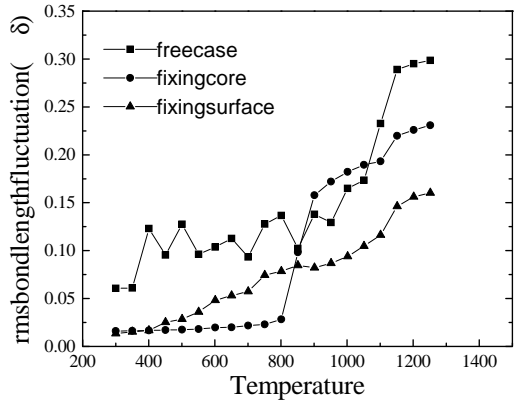}
}
\vspace{-0.7in}
\caption{Melting behavior in the 18-12-6-1 structure with fixing interior atoms and no constraint.}
\end{figure}

To check the validity of the current results, we have used the same MD code to study the melting of clusters and obtained results similar to previous Monte Carlo (MC) simulations\cite{wang1,wang2}. We have also exploited the melting process of gold nanowires by MC and observed similar interior diffuse behavior. Moreover, to examine the effect of the periodic bound condition on the melting behavior, we rescale the supercell length to 2-times and 3-times of original one. The melting temperatures are also about 1100K and the observed melting process is similar to the above results. It proves that the periodic bound condition has little effect on the simulation results. It should be further pointed out that we just limit our discussion to the helical structure in this paper. As mentioned above, this kind of helical structure is prevalent in metal nanowires in the small diameter range \cite{kondo,kondo1,tosatti,wang,bilalb1,bilalb2,bao}. Therefore, the current results on interior melting behavior in ultrathin gold nanowires is significant and might be a common feature in this kind of ultrathin metallic nanowires.

In summary, the thermal behavior of helical multi-walled gold nanowires has been studied and the main points are made as following. (1) The melting process starts from the interior region and no surface melting happens at lower temperature. We further argued that interior melting behavior happens prior to surface melting and surface melting represents the overall melting in ultrathin metallic nanowires. (2) The overall melting temperature of gold nanowires is lower than the one for the bulk, but higher than gold nanoclusters. (3) The surface and core atoms play different role in the melting behavior of these nanowires. The core atoms have a dominating effect on the melting at the beginning stage and the surface atoms are involved in the melting at higher temperature region. The core melting is closely related to the interior atomic structural characters. (4) The novel interior melting behavior is ultimately attributed to the helical multi-walled structure.

This work is financially supported by the National Natural Science Foundation of China(No.29890210, 10023001) and One-hundred-person project of Chinese Science Academic in China(2000).

\end{document}